\documentclass[onecolumn]{elsart3p}

\usepackage{amsmath}

\usepackage{pstcol}
\usepackage{color}

\usepackage{tikz}
\usetikzlibrary{decorations.pathmorphing,patterns}
\usetikzlibrary{calc,arrows}

\usepackage{graphicx}
 \usepackage{epsfig}

\usepackage{amssymb}
\usepackage[applemac]{inputenc}
\usepackage[T1]{fontenc}
\usepackage[english,francais]{babel}

\newtheorem{e-proposition}[theorem]{Proposition}

\newtheorem{e-definition}[theorem]{Definition\rm}

\renewcommand{\theequation}{\arabic{equation}}
\def\og{\leavevmode\raise.3ex\hbox{$\scriptscriptstyle\langle\!\langle$~}}
\def\fg{\leavevmode\raise.3ex\hbox{~$\!\scriptscriptstyle\,\rangle\!\rangle$}}

\definecolor{lgray}{gray}{0.9}
\definecolor{gray}{gray}{0.8}
\definecolor{dgray}{gray}{0.7}
\begin{document}
\centerline{Fourier and the science of today/\emph{Fourier et la sciences d'aujourd'hui}}
\begin{frontmatter}


\selectlanguage{english}
\title{Role of conserved quantities in Fourier’s law for diffusive mechanical systems}


\selectlanguage{english}
\author{Stefano Olla},
\ead{olla@ceremade.dauphine.fr}

\address{Ceremade, UMR CNRS, Universit\'e Paris Dauphine - PSL Research University\\
place Lattre de Tassigny, 75016 Paris Cedex 16, France}

\begin{abstract}
  Energy transport can be influenced by the presence of other conserved quantities.
  We consider here diffusive systems where energy and the other conserved quantities
  evolve macroscopically on the same diffusive space-time scale. In these situations the Fourier law
  depends also from the gradient of the other conserved quantities. The rotor chain is a classical example
  of such systems, where energy and angular momentum are conserved. 
  We review here some recent mathematical results about diffusive transport of energy and other
  conserved quantities, in particular for systems where the bulk Hamiltonian dynamics is perturbed by
  conservative stochastic  terms. The presence of the stochastic dynamics allows to define
  the transport coefficients (thermal conductivity) and in some cases to prove the local equilibrium
  and the linear response argument necessary to obtain the diffusive equations governing the macroscopic
  evolution of the conserved quantities. Temperature profiles and other conserved quantities profiles in the
  non-equilibrium stationary states can be then understood from the non-stationary diffusive behaviour.
  We also review some results and open problems on the two step approach
  (by weak coupling or kinetic limits) to the heat equation,
  starting from mechanical models with only energy conserved.
{\it To cite this article: S. Olla, C. R. Physique (2019).}

\vskip 0.5\baselineskip


 \vskip 0.5\baselineskip
}
\end{abstract}
\end{frontmatter}

\selectlanguage{english}
\section{Introduction}
\label{intro}

Fourier’s law claims that the local energy current is proportional to the local gradient of temperature
and the ratio of these quantities, which is a function of the local temperature,
is called the thermal conductivity.
However in many realistic systems energy is not the only conserved quantity
and the interplay between extra conserved quantities and energy has a deep impact
on the thermal properties of the system, for example generating uphill diffusion.

Dynamics that have more conserved quantities, either than energy,
often present different time scales for the macroscopic evolution of these.
In the present review we are interested in systems where conserved quantities evolve macroscopically
in the same \emph{diffusive} time scale, and their macroscopic evolution is governed by a system of
\emph{coupled} diffusive equations.
One example is given by the chain of coupled rotors,
whose dynamics conserves the energy and the angular momentum.
In Sections \ref{sec:line-resp-onsag} and \ref{sec:diff-macr-equat}
we show that, as a consequence of
a linear response argument and certain symmetries in the corresponding Onsager matrix,
the macroscopic evolution of these two quantities follows a diffusive system of PDE \eqref{eq:13}
or equivalently \eqref{eq:15}.
In terms of the evolution of the temperature profile, the usual heat diffusion can be
counterbalanced by a local increase of temperature due to the gradient of the
momentum profile (see \eqref{eq:16}).
It seems that there is some universality about equations \eqref{eq:16},
on the role of the gradients of the other conserved quantities. We give some examples where
the deterministic Hamiltonian dynamics is perturbed by a stochastic terms conservative of energy
and other quantities. In some of these stochastic models the hydrodynamic limit can be proven mathematically.

This macroscopic coupled evolution generates interesting stationary profiles of temperatures
in the non-equilibrium stationary states with thermal and mechanical forces
acting on the boundaries of the system,
as explained in section \ref{sec:non-equil-stat}.

In the last section we will report on some results concerning the two step approach for the Fourier law,
for dynamics that conserves only energy.

\section{Linear response and Onsager matrix }
\label{sec:line-resp-onsag}

For simplicity let us consider first a spatially homogeneous dynamics on a lattice $\mathbb Z$
such that there are two conserved quantities: \emph{momentum} and \emph{energy}.
A typical and commonly studied example is given by the rotor chain, where the configurations 
are given by $\{q_i(t)\in \mathbb S^1, p_i(t)\in \mathbb R, i\in \mathbb Z\}$,
where $\mathbb S^1$ is the unit circle. Nearest-neighbor rotators interact through a periodic potential
$V(q_{i+1} - q_i)$. For example $V(r) = 1- \cos(2\pi r)$, but the considerations
below are valid for more general interactions.
The Hamiltonian dynamics is given by
\begin{equation}
  \label{eq:2}
    \dot q_i(t) = p_i(t) , \qquad \dot p_i (t) = V'(q_{i+1}(t) - q_i(t)) -  V'(q_{i}(t) - q_{i-1}(t)).
  \end{equation}
  In the following we will denote $r_i = q_i - q_{i-1}$.
There are two locally conserved (or balanced) quantities: the momentum $p_i$
and the total energy $e_i= \frac{p_i^2}{2} + V(r_i)$. In fact we have
\begin{equation}
  \label{eq:1}
  \begin{split}
    \frac d{dt} p_i(t) &= j^p_{i-1,i}(t) - j^p_{i,i+1}(t), \qquad j^p_{i,i+1}(t) = -V'(r_{i+1}(t)),\\
    \frac d{dt} e_i(t) &= j^e_{i-1,i}(t) - j^e_{i,i+1}(t), \qquad j^e_{i,i+1}(t) = -p_i(t) V'(r_{i+1}(t)),
\end{split}
\end{equation}
where $j^\alpha_{i,i+1}(t), \alpha =p,e$,  are the corresponding instantaneous currents.
Notice that $r_i$  is \textbf{not} a balanced quantity.

One of the main conditions needed in order to have a macroscopic autonomous diffusive evolution
of these conserved quantities is that there are no other \emph{translation invariant}
conserved quantities for the dynamics of the \emph{infinite} system.
This \emph{ergodic} property is very hard to prove for the deterministic dynamics.
We may consider also some stochastic perturbation of the dynamics \eqref{eq:2} that have the same
conserved quantities. The purpose of such perturbations is in fact to guarantee that there are no other
balanced quantities \cite{ffl}.
One example is to add a random flip of the signs of $r_i$: for each particle $i$, at exponential
times with rate $\gamma$,  $r_i$ changes to $-r_i$, independently from the other particles $j\neq i$.
This operation does not change the momentum
.

A precise way to state this ergodic property of the infinite dynamics is the characterization of the
stationary and translation invariant probability measures, i.e. that the Gibbs measures
\begin{equation}
  \label{eq:gibbs}
  d\nu_{\beta, p} = \prod_{i\in\mathbb Z} \frac{e^{-\beta e_i + \beta p p_i}}{Z_{\beta,\beta p}} dp_i dr_i,
  \qquad \beta>0, \quad p\in \mathbb R, 
\end{equation}
are the only stationary and translation invariant probability measures for the dynamics
(within a certain class of \emph{regular} distributions).
We will denote by $<\cdot>_{\beta,p}$ the expectation with respect to $\nu_{\beta, p}$,
as well as the expectation of a function of the path of the dynamics under $\nu_{\beta, p}$.
These Gibbs measures are called \emph{equilibrium} and they have the following time-reversible property:
if $\{\left(\mathbf{r}(t), \mathbf{p}(t)\right) = \left(r_i(t), p_i(t), i\in \mathbb Z\right), t\in [-t_0, t_0]\}$
are stationary distributed with marginal $\nu_{\beta, p}$,
then $\{\tilde r_i(t) = r_i(-t), \tilde p_i(t) = - p_i(-t)\}$
follow the same dynamics, but with marginal $\nu_{\beta, -p}$,
i.e. for any function of the path $F(\mathbf r, \mathbf p)$ we have
$<F(\mathbf{\tilde r}(\cdot), \mathbf{\tilde p}(\cdot))>_{\beta,p} = <F(\mathbf r(\cdot), \mathbf p(\cdot))>_{\beta,-p}$.

In the specific case of the rotators, there is also a \emph{rotational symmetry} of the equilibrium dynamics, i.e.
under the equilibrium $(\beta, p)$, the distribution of the path of
$\tilde r_i(t) = r_i(t), \tilde p_i(t) = p_i(t) - p$
is the same as the one at equilibrium $(\beta, 0)$, i.e. with zero average velocity.

One way to define the transport diffusion coefficients is through \emph{linear response}.
Notice that for any equilibrium state, we have that $<j_{i,i+1}^p>_{\beta,p} = 0 = <j_{i,i+1}^e>_{\beta,p}$ for any values of $\beta>0$ and $p\in \mathbb R$,
consequently we do not expect any ballistic evolution in these systems. 
We have to understand how these expectations behave if we impose a gradient of temperature or of
momentum, at the first order for small gradients.
Given $\epsilon_1, \epsilon_2$, 
consider the inhomogeneous Gibbs measure
\begin{equation}
  \label{eq:gibbs-grad}
  d\mu_{\epsilon_1,\epsilon_2} = \prod_{i\in\mathbb Z}
  \frac{e^{-(\beta + \epsilon_1 i) e_i + (\beta  p +\epsilon_2 i) p_i}}{Z_{\beta+\epsilon_1 i,\beta p + \epsilon_2 i}}
  dp_i dr_i.
\end{equation}
Starting at time $t=0$ with this inhomogeneous measure we expect that, at the first order in
$(\epsilon_1, \epsilon_2)$,
\begin{equation}
  \label{eq:linres}
  \begin{split}
    <j_{0,1}^p(t) >_{\mu_{\epsilon_1,\epsilon_2} } &=
    K^{p,p}_t \epsilon_2 + K^{p,\beta}_t \epsilon_1 + o(\epsilon_1,\epsilon_2)\\
    <j_{0,1}^e(t) >_{\mu_{\epsilon_1,\epsilon_2} } &=
    K^{\beta,p}_t \epsilon_2 + K^{\beta,\beta}_t \epsilon_1 + o(\epsilon_1,\epsilon_2),
  \end{split}
\end{equation}
and then we are interested in the limit as $t\to\infty$ for the coefficients $K^{u,v}_t$.
Defining $e =  \left<e_i \right>_{\beta, p}$, a straightforward development at the first order gives
\begin{equation}
  \label{eq:5}
  \begin{split}
  K^{p,p}_t &= \sum_i i \left< j_{0,1}^p(t) (p_i(0)- p) \right>_{\beta, p},\quad
  K^{p,\beta}_t = -\sum_i i \left< j_{0,1}^p(t) (e_i(0)- e) \right>_{\beta, p}, \\
   K^{\beta,p}_t &= \sum_i i \left< j_{0,1}^e(t) (p_i(0)- p) \right>_{\beta, p},\quad
  K^{\beta,\beta}_t = -\sum_i i \left< j_{0,1}^e(t) (e_i(0)- e) \right>_{\beta, p}, 
\end{split}
\end{equation}
assuming that the sums in \eqref{eq:5} converge.
By using the symmetries of the dynamics
(rotational, time stationarity, time reversibility), and recalling that
$ \left< j_{0,1}^p(0) p_i(0) \right>_{\beta, 0} = 0$,  we have
\begin{equation}
  \label{eq:6}
  \begin{split}
  K^{p,p}_t (\beta, p) &= \sum_i i \left< j_{0,1}^p(t) p_i(0) \right>_{\beta, 0} =
  - \sum_i i \left< j_{0,1}^p(0) p_i(t) \right>_{\beta, 0}\\
  &= - \int_0^t ds \sum_i i \left< j_{0,1}^p(0) \left( j_{i-1, i}^p(s) -  j_{i, i+1}^p(s)\right) \right>_{\beta, 0}
  = - \int_0^t ds \sum_i  \left< j_{0,1}^p(0)  j_{i,i+1}^p(s) \right>_{\beta, 0}.
\end{split}
\end{equation}
So we define the limit as $t\to\infty$, assuming that it exists, as
\begin{equation}
  \label{eq:7}
  K^{p,p}(\beta,p) =  - \int_0^\infty ds \sum_i  \left< j_{0,1}^p(0)  j_{i,i+1}^p(s) \right>_{\beta, 0} = K^{p,p}(\beta,0)
  : =  K^{p,p}(\beta). 
 \end{equation}
Notice that $K^{p,p}$ is only a function of $\beta$ and does not depend on $p$.
This is a consequence of the rotational symmetry of the dynamics.
We define similarly $ K^{p,\beta}(\beta,p),  K^{\beta,p}(\beta,p),  K^{\beta,\beta}(\beta,p)$.

Similar calculations give, recalling that $\left< j_{0,1}^p(0) e_i(0) \right>_{\beta,  p} = 0$
for any $p$ and $\beta$,
\begin{equation}
  \label{eq:8}
  \begin{split}
    K^{p,\beta}_t =  - \sum_i i \left< j_{0,1}^p(t) (e_i(0)- e) \right>_{\beta, p}
    = - \sum_i i \left< j_{0,1}^p(0) (e_i(t)- e) \right>_{\beta, - p}
    = - \int_0^t ds  \sum_i \left< j_{0,1}^p(0) j^e_{i,i+1} (s) \right>_{\beta, - p}\\
    = - \int_0^t ds \sum_i \left< j_{0,1}^p(0) p_i(s) j^p_{i+1,i+2} (s) \right>_{\beta, - p}
    = \int_0^t ds \sum_i \left< j_{0,1}^p(s) p_i(0) j^p_{i+1,i+2} (0) \right>_{\beta, p}\\
    = - p K^{pp}_t + \int_0^t ds \sum_i \left< j_{0,1}^p(s) (p_i(0)- p) j^p_{i+1,i+2} (0) \right>_{\beta, p}
     = - p K^{pp}_t + \int_0^t ds \sum_i \left< j_{0,1}^p(s) p_i(0) j^p_{i+1,i+2} (0) \right>_{\beta, 0}
  \end{split}
\end{equation}
The second term on the right hand side is equal to $0$, since $\int_0^t ds \sum_i  j_{0,1}^p(s)$
is symmetric by time reversal $s \to t-s$
while $p_i(0) j^p_{i+1,i+2} (0)$ is antisymmetric. This implies that $ K^{p,\beta} = - p K^{pp}$.
A similar argument gives 
\begin{equation}
K^{\beta,p}(\beta, p) =  K^{p,\beta}(\beta, -p) =  p K^{pp}(\beta). \label{eq:11}
\end{equation}

Similarly, 
\begin{equation}
  \label{eq:10}
  \begin{split}
    K^{\beta,\beta}_t = - \sum_i i \left< j_{0,1}^e(t) (e_i(0)- e) \right>_{\beta, p} =
     \sum_i i \left< j_{0,1}^e(0) (e_i(t) - e_i(0)) \right>_{\beta, -p} \\
    =   \int_0^t ds \sum_i \left< j_{0,1}^e(0) j^e_{i,i+1}(s) \right>_{\beta, -p} 
    =  \int_0^t ds \sum_i \left< j_{0,1}^e(0) j^e_{i,i+1}(s) \right>_{\beta, p},
 \end{split}
\end{equation}
Using again the symmetries and recalling that $ j^e_{i,i+1} = p_i j^p_{i+1,i+2}$ we obtain also that
\begin{equation}
  \label{eq:9}
   K^{\beta,\beta}(\beta, p) = K^{\beta,\beta}(\beta, 0) - p^2 K^{p,p}(\beta).
\end{equation}
Finally thanks to the symmetries of the system,
all the coefficients can be calculated from $ K^{p,p}(\beta)$ and
$K^{\beta,\beta}(\beta) := K^{\beta,\beta}(\beta,0)$, computed at zero average velocity.
Relations \eqref{eq:11} and \eqref{eq:9} were already noted in \cite{illp16}. 

One of the main mathematical problems in dealing with the deterministic infinite dynamics,
is in proving that the limits defining
$ K^{p,p}(\beta)$ and $K^{\beta,\beta}(\beta)$ exist and are finite. If this is an open problem
for the deterministic dynamics, stochastic perturbations can help.
In fact adding a random independent flip of the $r_i$'s one can prove that
$ K^{p,p}(\beta)$ and $K^{\beta,\beta}(\beta)$ are well defined and finite, by adapting the argument used in
\cite{bo2}. 

\section{The diffusive macroscopic equations}
\label{sec:diff-macr-equat}

\subsection{Macroscopic diffusive equations for the rotors model}
\label{sec:macr-diff-rotor}

The linear response analysis of the previous section gives a heuristic argument for the macroscopic
equations governing the evolution in the diffusive space--time scale.
In order to state the macroscopic equations, we need some thermodynamic functions.
The \emph{internal energy} (or \emph{thermal energy}) as a function of $\beta$ is given by
$u(\beta) = -\partial_\beta \log Z_{\beta,0}$, where $Z_{\beta,0}$ is the partition function appearing in
\eqref{eq:gibbs}.
The temperature is given by $T = \beta^{-1}$, and the heat capacity
is defined as $c_v(T) = \frac{du}{dT} = \beta^2 \text{Var}_{\beta, 0}(e_i)$. 
The thermodynamic entropy is $S(u) = \inf_{\beta>0} \left\{ \beta u + \log Z_{\beta,0}\right\}$,
and $\beta(u) = S'(u)$ provides the inverse function of $u(\beta)$.

The linear response argument \eqref{eq:linres} suggests the following macroscopic equations
for the evolution of the profiles $p(t,x), e(t,x)$ of the conserved quantities:
\begin{equation}
  \label{eq:12}
  \begin{split}
    \partial_t p &= -\partial_x \left(K^{p,p}(\beta) \partial_x (\beta p) + K^{p,\beta}(\beta,p) \partial_x \beta\right)\\
    \partial_t e &= -\partial_x \left(K^{\beta,p}(\beta,p) \partial_x (\beta p)
      + K^{\beta,\beta}(\beta,p) \partial_x \beta\right),
  \end{split}
\end{equation}
with the profiles $\beta(t,x) := \beta(u(t,x))$, and $u(t,x) = e(t,x) - \frac 12 p^2(t,x)$.
By using the relations \eqref{eq:11} and \eqref{eq:9},
the only coefficient involved are $K^{p,p}(\beta)$ and
$K^{\beta,\beta}(\beta) =K^{\beta,\beta}(\beta,0)$.
Expressing the equations in terms of the temperature profile
$T(t,x) = \beta^{-1}(t,y)$, \eqref{eq:12} are equivalent to
\begin{equation}
  \label{eq:13}
   \begin{split}
    \partial_t p &= \partial_x \left(D^p(T) \partial_x p\right)\\
    \partial_t e &= \partial_x \left(D^p(T) \partial_x\left(\frac{p^2}2\right)
      + \kappa(T) \partial_x T \right),
  \end{split}
\end{equation}
where
\begin{equation}
  \label{eq:14}
  \begin{split}
  D^p(T) &:=  -T^{-1} K^{p,p}(T^{-1})  \qquad \text{momentum diffusivity},\\
  \kappa (T) &: = T^{-2} K^{\beta,\beta}(T^{-1}) \qquad\quad  \text{thermal conductivity}.
\end{split}
\end{equation}
Alternatively, rewriting \eqref{eq:13} as closed equations in $p(t,x)$ and $e(t,x)$:
\begin{equation}
  \label{eq:15}
  \begin{split}
    \partial_t p &= \partial_x \left(D^p \partial_x p\right)\\
    \partial_t e &= \partial_x \left((D^p - D^e)\partial_x\left(\frac{p^2}2\right)
      + D^e \partial_x e\right),
  \end{split}
\end{equation}
with the energy diffusivity defined by $D^e = \frac{\kappa(T)}{c_v(T)}$.

It is more interesting to close the equations in the temperature profile $T(t,x)$ obtaining
\begin{equation}  \label{eq:16}
   \begin{split}
    \partial_t p &= \partial_x \left(D^p(T) \partial_x p\right)\\
    c_v(T) \partial_t T &= \partial_x\left(\kappa(T) \partial_x T\right) + D^p(T) (\partial_x p)^2 .
  \end{split}
\end{equation}
There are two remarkable consequences from the equations \eqref{eq:16}:
\begin{itemize}
\item The gradient in the temperature does not contribute to the diffusion of the  momentum,
  but the momentum diffusivity depends only on the temperature.

\item The gradient of the momentum \textbf{increases} locally the temperature.
  The time evolution of the temperature is composed by the usual heat diffusion term
  $\partial_x\left(\kappa(T) \partial_x T\right)$
  plus an increase due to inhomogeneities in the momentum distribution. 
\end{itemize}

The total energy $e(t,x)$ is the sum of the mechanical energy $p^2(t,x)/2$ and an \emph{internal}
energy $u(t,x)$.
Notice that $\partial_t u =  c_v(T) \partial_t T$, i.e.
the momentum diffusion decreases the mechanical energy, that dissipates
into internal energy, and
the term $D^p(T) (\partial_x p)^2$ is the rate of dissipation of the mechanical energy in
internal energy. We can also read this in the increase of the total entropy of the system:
\begin{equation}
  \label{eq:24}
  \frac d{dt} \int S(u(t,x)) dx = \int \left[\frac{D^p(T)}{T} (\partial_x p)^2
    + \frac{\kappa(T)}{T^2} (\partial_x T)^2\right] dx.
\end{equation}

\subsection{Mathematical problems }
\label{sec:math-probl-}

We have obtained the equations \eqref{eq:16} heuristically from the linear response argument and
some symmetries of the dynamics.
In fact they are quite general, the details of the microscopic dynamics
are contained in the macroscopic transport coefficients $D^p(T)$ and $\kappa(T)$.
A rigorous mathematical statement would be given by a \emph{hydrodynamic limit}:
consider the empirical profile distributions
\begin{equation}
  \label{eq:17}
  \hat p_\epsilon(t,x) = \epsilon \sum_i p_i(\epsilon^{-2}t) \delta_{\epsilon i}(x), \qquad
   \hat e_\epsilon(t,x) = \epsilon \sum_i e_i(\epsilon^{-2}t) \delta_{\epsilon i}(x).
 \end{equation}
 These are random variables valued on distributions on $\mathbb R$. We would like to prove that,
 as $\epsilon\to 0$, they converge in probability to the solution $p(t,x), e(t,x)$ of \eqref{eq:15}. 
 We have already mentioned that for the deterministic dynamics, even the existence of
 $D^p(T)$ and $\kappa(T)$ is an open problem. Adding stochastic conservative terms to the dynamics
 can help to prove the existence of the transport coefficient \cite{bo2},
 but still the proof of the hydrodynamic limit is a difficult task.
 The main problem is the following.
 In the hydrodynamic limit, on the microscopic scale, gradients have order
 $\epsilon$, but one has to look at a time scale $\epsilon^{-2}$.
 In the linear response argument we have
 first made an expansion in the first order of the gradients (cf. \eqref{eq:linres}) because of a large space scale,
 and subsequently we took the large time limit.
 In the hydrodynamic limit we have to take the limit in space and time together, with the diffusive scaling.

 It should be mentioned here that this possible \emph{mathematical} statement about the diffusive limit
 does not explain the superdiffusive behavior that the rotor chain may have at low temperatures
 in certain space-time scales. This is due to a kind of \emph{metastable} low temperature states,
 where rotors have mostly an oscillatory behaviour, like an FPU chain of anharmonic springs,
 and the $r_i$ behave like an \emph{almost conserved quantity} (see discussion and simulations in
 \cite{spohn-rot,das-dhar} and for a related model in \cite{ddd}).

 A technique to approach this hydrodynamic limit problem in stochastic dynamics
 was developped by Varadhan \cite{Var}
 (see also chapter 7 in \cite{kipnis}). It consists in decomposing, by approximations, the microscopic
 currents in a gradient term (dissipation) plus a term in the range
 of the generator $L$ of the dynamics (fluctuation).
 In the present context this would mean 
 \begin{equation}
   \label{eq:19}
   j^{p}_{0,1}\ \sim\  -D^{p}(T) (p_1-p_0) + LF^p, \qquad
   j^{e}_{0,1}\ \sim\ - D^e(T) (e_1 - e_0) - p(D^p(T) - D^e(T)) (p_1 - p_0) + LF^e,
 \end{equation}
 where the approximation symbol means that there exist sequences of local functions $F^p, F^e$
 in the domain of the generator $L$ such that the space-time variance of the difference,
 calculated for the
 dynamics in the equilibrium with average momentum $p$ and temperature $T$, vanishes.
 Such fluctuation-dissipation decompositions allow to substitute, locally, \emph{currents}
 with \emph{gradients}
 and eventually close the equations after the hydrodynamic limit.
 In \cite{Var} Varadhan used first this approach to prove the hydrodynamic limit
 for a non-gradient reversible stochastic
 Ginzburg-Landau dynamics (one conserved quantity).
 Reversibility here is intended for a dynamics that has a self-adjoint
 generator with respect to the stationary Gibbs measures.
 Then Quastel, in his PhD thesis \cite{quastel}, proved by this technique the diffusion of colors in the
 symmetric simple exclusion process (still reversible but with two conserved quantities).
 There have been many other results extending this method to non-reversible dynamics.
 In the context  of a chain of anharmonic oscillators with a stochastic perturbation
 conserving only the energy,
 the fluctuation-dissipation decomposition has been proved in \cite{os}.
 One of the limitations of this non-gradient approach is that it requires that the generator of the
 dynamics $L$ has a very 'large' symmetric part $S$, whose finite dimensional version has a
 spectral gap uniformly bounded with respect to the dimension;
 furthermore $L$ should satisfy a spectral sector
 condition with respect to $S$. This requires to consider only stochastic perturbations
 that act on both positions and velocities  ($S$ locally elliptic).

 \subsection{Macroscopic diffusive equations for the harmonic chain with bulk noise}
\label{sec:macr-diff-equat-1}

 In the case when $\kappa$ and $D^e$ are constant independent of the temperature $T$,
 there are examples of dynamics where the hydrodynamic limit can be proven rigorously.
 This is the case of a chain of harmonic oscillators whose Hamiltonian dynamics is perturbed by
 a random sign change of the interparticle distance.
 The dynamics is defined as in \eqref{eq:2}, but now $r_i\in \mathbb R$ and $V(r) = \frac{r^2}2$.
 Furthermore any particle $i$ has an independent Poisson process with intensity $\gamma$,
 when it rings $r_i$ changes sign. Momentum and energy are conserved and it can be proven that
 the empirical distribution defined by \eqref{eq:17} converges to the solution of \eqref{eq:15},
 with explicit $\kappa$ and $D^e$ depending only on $\gamma$. Of course these coefficients diverge
 for $\gamma\to 0$ as the deterministic harmonic chain does not have a diffusive behaviour.
 
 A rigorous mathematical proof of such hydrodynamic limit for a different dynamics of springs,
with random perturbations that  do not conserve momentum but conserves $r_i$
 can be found in \cite{ber05,kos1}. In \cite{kos1} is considered a harmonic chain,
 but with a random flip of the velocities sign.
 In this case the two conserved quantities are the volume stretch $r_i$ (with currents $j^r_{i,i+1} = -p_i$)
 and the energy. In \cite{kos1} it is proven that the empirical distributions
 \begin{equation}
  \label{eq:17}
  \hat r_\epsilon(t,x) = \epsilon \sum_i r_i(\epsilon^{-2}t) \delta_{\epsilon i}(x), \qquad
   \hat T_\epsilon(t,x) = \epsilon \sum_i \frac 12 p^2_i(\epsilon^{-2}t) \delta_{\epsilon i}(x).
 \end{equation}
 converge to the solution $r(t,x), T(t,x)$ of
 \begin{equation}
   \label{eq:20}
   \begin{split}
    \partial_t r &= \frac {1}{2\gamma}  \partial^2_x r \\
    \partial_t T &= \frac {1}{2\gamma}  (\partial_x r)^2 + \frac {1}{4\gamma} \partial^2_x T,
  \end{split}
 \end{equation}
 that have the same structure as \eqref{eq:16}.
 Similar results with a different stochastic perturbation are contained in \cite{ber05}.
 In these models,
 where the transport coefficients are constant,  the fluctuation-dissipation relations \eqref{eq:19}
 are exact for explicit local functions $F$. Still, the equations \eqref{eq:20} are non linear and the proof
 in \cite{kos1} demands the application of Wigner distributions techniques in order to control
 the separate evolution of the thermal and mechanical energy.

 For the anharmonic chain with velocity flip, the corresponding macroscopic equations are given by
 \cite{so1,so2}
 \begin{equation}
   \label{eq:40}
   \begin{split}
    \partial_t r &= \frac {1}{2\gamma}  \partial^2_x \tau(r,T) \\
    c_v(r,T) \partial_t T &= \partial_x\left(\kappa(r,T) \partial_x T\right) +  \frac 1{2\gamma} (\partial_x \tau(r,T))^2 ,
  \end{split}
 \end{equation}
 where $\tau(r,T)$ is the thermodynamic tension at temperature $T$ and volume $r$, $c_v(r,T) $
 the corresponding heat capacity. The thermal conductivity $\kappa(r,T)$ may depend on $r$ in a non explicit way,
 as we do not have the same symmetries as in the rotor model.
 Notice that in this case, thanks to the noise in the dynamics,
 the current of the volume strain is $j^r_{0,1} = -p_1$ and it has an explicit fluctuation--dissipation decomposition of the 
 \eqref{eq:19} type:
 \begin{equation}
   \label{eq:41}
   j^r_{0,1} = -p_1 = \frac1{2\gamma} Lp_1 - \frac1{2\gamma}\left(V'(r_1) - V'(r_0)\right), 
 \end{equation}
that explains the first of the equations \eqref{eq:40}.

 \subsection{Dynamics with 3 conserved quantities}
\label{sec:dynamics-with-3}

There are dynamics with 3 conserved quantities that evolve macroscopically
in the diffusive space-time scaling. In \cite{kona}
we study a harmonic chain where the potential energy
does not depend on the volume strain or the tension of the system, but on its bending or \emph{curvature}.
This implies that we add springs between next nearest neighbor particles with negative potential,
such that the total potential energy is given by
\begin{equation}
  \label{eq:21}
  \frac 12 \sum_i \left(q_{i+1} - q_{i-1} - 2 q_i\right)^2.
\end{equation}
To the Hamitonian dynamics we add a random exchange, with intensity $\gamma$,
of velocities between nearest neighbor particles.
It results that there are three conserved quantities:
\begin{equation}
  \label{eq:22}
  \begin{split}
    k_i = q_{i+1} - q_{i-1} - 2 q_i, 
    \qquad p_i ,\qquad
    e_i = \frac{p_i^2}2 + \frac{k_i^2}{2}
\end{split}
\end{equation}
We call  $k_i$ the \emph{curvature or bending}. In \cite{kona}
we prove that the corresponding empirical distributions converge to
$k(t,x), p(t,x), e(t,x)$ solutions of the diffusive system
\begin{equation}
  \label{eq:23}
  \begin{split}
    \partial_t k &= -\partial^2_x p\\
    \partial_t p &= \partial^2_x k + \gamma \partial^2_x p\\
    \partial_t T &= \frac{1}{\gamma} \partial^2_x T + \gamma \left(\partial_x p\right)^2,
  \end{split}
\end{equation}
where the temperature profile is defined as $T(t,x) = e(t,x) - \frac{p^2(t,x)}{2}$.
We can see this as a diffusive perturbation of the Bernoulli's beam wave equation
$\partial_t^2 k = -\partial_x^4 k$.
Notice the similar structure as in equations \eqref{eq:16} and \eqref{eq:20}, with a \emph{heating}
term $\gamma \left(\partial_x p\right)^2$.

It is an open question if similar macroscopic equations will hold for a non linear dynamics with potential
$V(k_i)$, without any stochastic term.
Numerical dynamical simulations have been inconclusive about this possible diffusive behaviour.

\section{Non-equilibrium stationary states}
\label{sec:non-equil-stat}

\subsection{Stationary temperature profiles and up-hill diffusion in rotor model}
\label{sec:stat-temp-prof}

In this section we review some results concerning the non-equilibrium stationary states for the rotors model
contained in \cite{iaco2}, more details can be found there.
Let us consider the finite dynamics on $N+1$-rotors as in \eqref{eq:2},
where we add boundary forces or heath baths
such that gradients are imposed on the conserved quantities
in the corresponding stationary state.
In order to establish a gradient in the temperature profile, we apply at the boundary two Langevin
heat baths at different temperatures $T_L, T_R$,  while constant forces $\tau_L, \tau_R$
act respectively on the first rotor on the LHS and last rotor on the RHS.
The equations of motion read as:
\begin{equation}
  \label{eq:2ss}
  \begin{split}
    \dot r_i(t) &= p_i(t) - p_{i-1}(t) , \qquad 1, \dots , N, \\
    \dot p_i (t) &= V'(r_{i+1}(t)) -  V'(r_{i}(t)), \quad i=2, \dots, N-1,\\
     dp_0 (t) &= \left(\tau_L + V'(r_{1}(t)) - \gamma p_0(t)\right) dt + \sqrt{2\gamma T_L} dw_L(t),\\
     dp_N (t) &= \left(\tau_R -  V'(r_{N}(t)) - \gamma p_N(t) \right) dt + \sqrt{2\gamma T_R} dw_R(t),
  \end{split}
\end{equation}
where $w_L(t), w_R(t)$ are two independent Wiener processes.

We expect here that an hydrodynamic limit holds for the empirical profile distribution \eqref{eq:19}
  scaling with $\epsilon = N^{-1}$, with the equation \eqref{eq:16} in $x\in [0,1]$ provided with the
  boundary conditions
  $$p(t,0) = \frac{\tau_L}{\gamma},\quad  T(t,0) = T_L, \qquad p(t,1) = \frac{\tau_R}{\gamma},
  \quad T(t,1) = T_R.$$

  As $t\to\infty$ the system, at fixed size $N$, should approach a stationary state $\mu^N_{ss}$
  depending in principle on $T_L, T_R, \tau_L, \tau_R$ and $\gamma$.
  This probability distribution is called
  non-equilibrium stationary state (NESS) and only in the case $T_L = T_R = \beta^{-1}$
  and
  $\tau_L = \tau_R = \tau$ it coincides with the equilibrium measure $\nu_{\beta,p}$ with $p = \gamma^{-1} \tau$.


  From the mathematical side, the study of the NESS and its asymptotic properties as $N\to\infty$
  is harder than the non stationary behavior. This is because in the NESS the time scales are hidden.
  For the rotors model, even the existence
  of the NESS is an open problem, and only recently there have been
  some progress for $N=2$ and $3$  \cite{cuneo1,cuneo2}.

  Assuming the existence of the NESS, by stationarity the expectation of the
  currents of the conserved quantities have to be homogeneous along the chain,
  i.e. denoting with $<\cdot>_{ss}$ the expectation with respect to $\mu^N_{ss}$, we have
  \begin{equation}
    J_N^p := < j^p_{i,i+1}>_{ss} , \qquad J^e_N:= <j^e_{i,i+1}>_{ss} , \qquad i=1, \dots, N-1,
  \end{equation}
  are constant in $i$.  Taking into account also the boundary currents we have
  \begin{equation}
    \label{eq:25}
    \begin{split}
      J_N^p\ &=\ -\tau_R + \gamma <p_N>_{ss}\  =\ \tau_L - \gamma <p_0>_{ss}, \\
      J_N^e\ &= \ \gamma\left(T_L - <p_0^2>_{ss} \right) -\tau_L <p_0>_{ss}\
      = \ \gamma\left( <p_N^2>_{ss} - T_R \right) + \tau_R <p_N>_{ss}.
  \end{split}
  \end{equation}
  The diffusive behaviour implies that $J_N^p, J_N^e \sim O\left(\frac 1N\right)$, and we expect that
  \begin{equation}
    \label{eq:3}
    \begin{split}
      \lim_{N\to\infty} N J_N^p &= J^p = -D^p(T_{ss}(x)) \partial_x p_{ss}(x)\\
      \lim_{N\to\infty} N J_N^e &= J^e = -D^p(T_{ss}(x)) \partial_x\left(\frac{p_{ss}(x)^2}2\right)
      - \kappa(T_{ss}(x)) \partial_x T_{ss}(x),
    \end{split}
  \end{equation}
  where $p_{ss}(x) ,T_{ss}(x)$ are the stationary solutions of equations \eqref{eq:13} with boundary conditions
  $$p_{ss}(0) =  \frac{\tau_L}{\gamma},\qquad T_{ss}(0) = T_L,\qquad  p_{ss}(1) = \frac{\tau_R}{\gamma} ,\qquad T_{ss}(1) = T_R.$$

  
  Notice that the energy current is the sum of the \emph{heat current}
  $J^Q(x) = - \kappa(T_{ss}(x)) \partial_x T_{ss}(x)$, and the \emph{mechanical energy current}
  $ -D^p(T_{ss}) \partial_x\left(\frac{p_{ss}^2}2\right)$. They can be of opposite signs, giving rise to the
  phenomenon of \emph{uphill diffusion} ($J^e$ of the same sign as
  the gradient of temperature, cf. \cite{krihsna}).

  Some other relations can be obtained from \eqref{eq:3}. By multiplying the first equation by $p_{ss}(x)$
  and subtracting the second we have
  \begin{equation}
    \label{eq:18}
    p_{ss}(x) J^p - J^e = \kappa(T_{ss}(x)) \partial_x T_{ss}(x) = - J^Q(x),
  \end{equation}
 while the second derivative of $T_{ss}(x)$ must satisfy
  \begin{equation}\label{eq:18b}
    J^p \partial_x p_{ss}(x) = \kappa(T_{ss}(x)) \partial_{xx}T_{ss}(x)
    + \kappa'(T_{ss}(x)) \left(\partial_{x}T_{ss}(x)\right)^2 .
  \end{equation}


  Equations \eqref{eq:3} predict a maximum for the temperature profile $T_{ss}(x)$ inside the interval
  $(0,1)$ higher than the boundary temperatures $T_L, T_R$. In fact, without losing generality,
  assume $\tau_R = \tau >0$ and $\tau_L = 0$, then $p_{ss}(x) \ge 0, J^p <0$
  and $p_{ss}(x)$ is strictly increasing from $p_{ss}(0) = 0$ to $p_{ss}(1) = \tau/\gamma$.
  Consequently, from \eqref{eq:18}, we can have only one stationary point for $T_{ss}$, and by
   \eqref{eq:18b} it must be a maximum, that we denote by $x_{max}$, that
  must satisfy $p_{ss}(x_{max}) J^p = J^e$.
  This implies that $J^p$ and $J^e$ are of the same sign and,
  if $J^e\neq 0$, the strict increase property of  $p_{ss}$ implies that
  $x_{max}$ must be inside the interval $(0,1)$. Since there are no other stationary points, the
  maximal temperature $T_{ss}(x_{max})$ must be higher than the temperatures at the boundaries.

  From \eqref{eq:18b}, a flex point $x_{flex}$ of $T_{ss}(x)$ must satisfy the relation
  \begin{equation*}
    - D^p(T_{ss}(x_{flex}))  \left(\partial_x p_{ss}(x_{flex})\right)^2
    = J^p \partial_x p_{ss}(x_{flex}) = \kappa'(T_{ss}(x_{flex})) \left(\partial_{x}T_{ss}(x_{flex})\right)^2
  \end{equation*}
  It follows that such flex points can exist only around values where $\kappa(T)$
  is a strictly decreasing function of $T$.

 Another qualitative property of the solution can be seen from the first of \eqref{eq:3}. i.e. 
    we have that $\partial_x p_{ss}(x) = \frac{-J^p}{D^p(T_{ss}(x))}$. There is a numerical evidence that
    $D^p(T)$ is a decreasing function of $T$, so we have that $\partial_x p_{ss}(x)$ is proportional to
    $T_{ss}(x)^\alpha$ for some $\alpha$. It follows that
    \begin{equation}\label{eq:pflex}
      \partial_{x}^2 p_{ss} (x) = \frac{J^p (D^p)'(T_{ss}(x))}{D^p(T_{ss}(x))^2} \partial_x T_{ss}(x),
    \end{equation}
    that implies a flex point for $p_{ss}(x)$ at the point of maximum temperature.

    The energy current $J^e$ can present a negative linear response with respect to $T_R - T_L$,
    due to the decrease of $D^p(T)$ as a function of the temperature.
    In fact, in the case $\tau_R >0, \tau_L = 0$ and $T_R\ge T_L$,
    we have that $\partial_x\left(\frac{p_{ss}^2(x)}{2}\right) >0$, increasing $T_R$ will increase the whole profile
    $T_{ss}(x)$. This may create a positive increase to
    $- D^p(T_{ss}(x)) \partial_x\left(\frac{p_{ss}^2(x)}{2}\right)$ larger that the negative increase of
    $ J^q = - \kappa(T_{ss}(x)) \partial_x T_{ss}(x)$.


From the stationary equation we can compute the entropy production of the stationary state, that we can define as
    \begin{equation}
  \label{eq:30}
 \Sigma = \left(T_R^{-1} - T_L^{-1}\right) J^e
  - \gamma^{-1} \left(T_R^{-1}\tau_R - T_L^{-1}\tau_L\right) J^p
\end{equation}
It turns out that $\Sigma >0$ and that it is equal to
\begin{equation}
  \label{eq:31}
  \Sigma = \int_{-1}^{1} \left[\frac{\kappa(T_{ss}(x))}{T_{ss}^2(x)} \left(\partial_x T_{ss}(x)\right)^2
  + \frac{D^p(T_{ss}(x)) }{T_{ss}(x)} \left(\partial_x p_{ss}(x)\right)^2 \right] dx.
\end{equation}
    Notice that this expression coincide with the time derivative of Clausius entropy $S$ given in \eqref{eq:24}.

\begin{figure}
    \centering
     \includegraphics[width=0.70\textwidth]{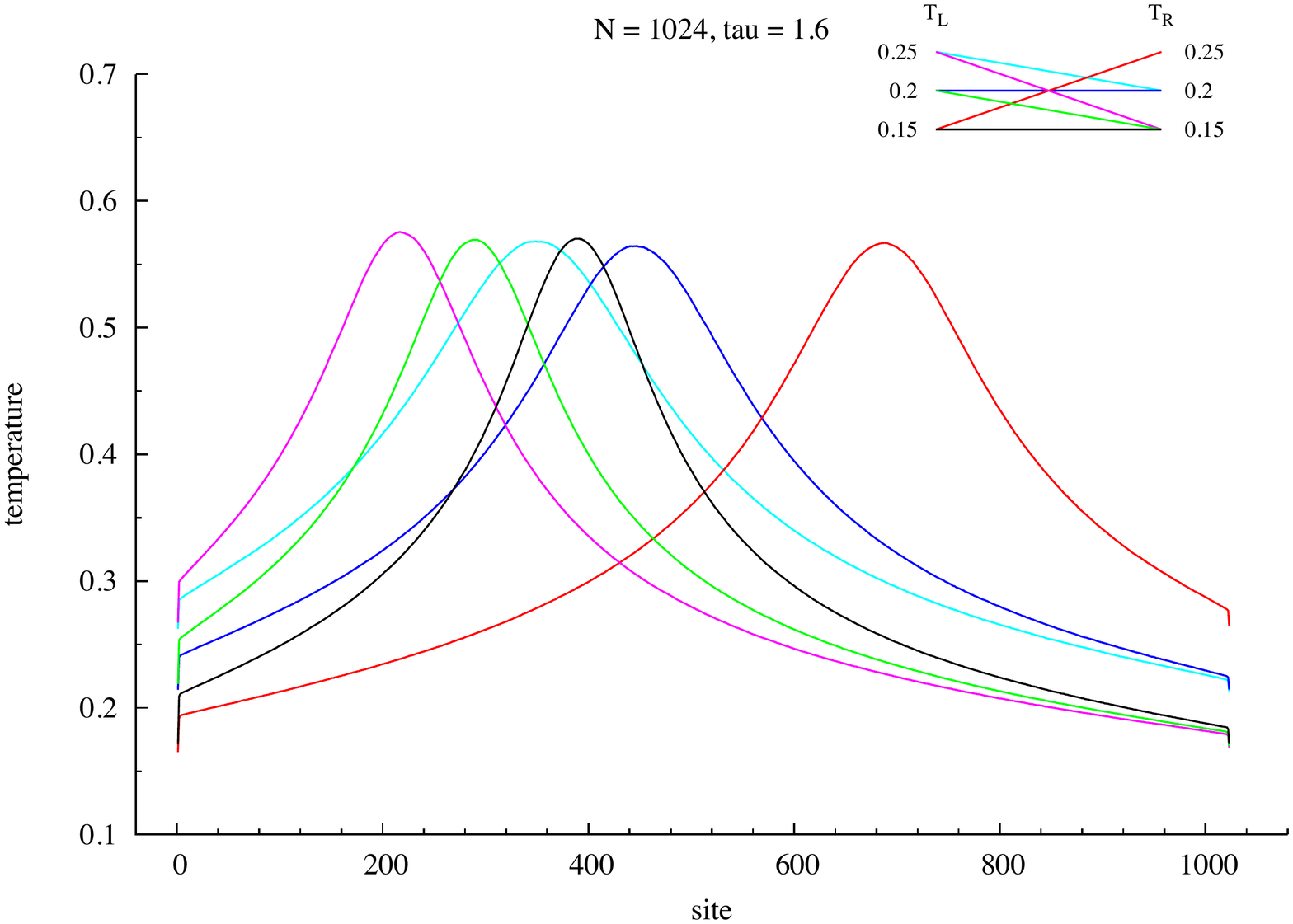}
     \caption{Stationary temperature profiles in rotors dynamics, with $\tau_R = \tau, \tau_L = 0$
       and different values of temperatures for the Langevin heat bath. Reprinted from \cite{iaco}.}
    \label{fig:1}
  \end{figure}

\begin{figure}
 \centering
  \includegraphics[width=0.70\textwidth]{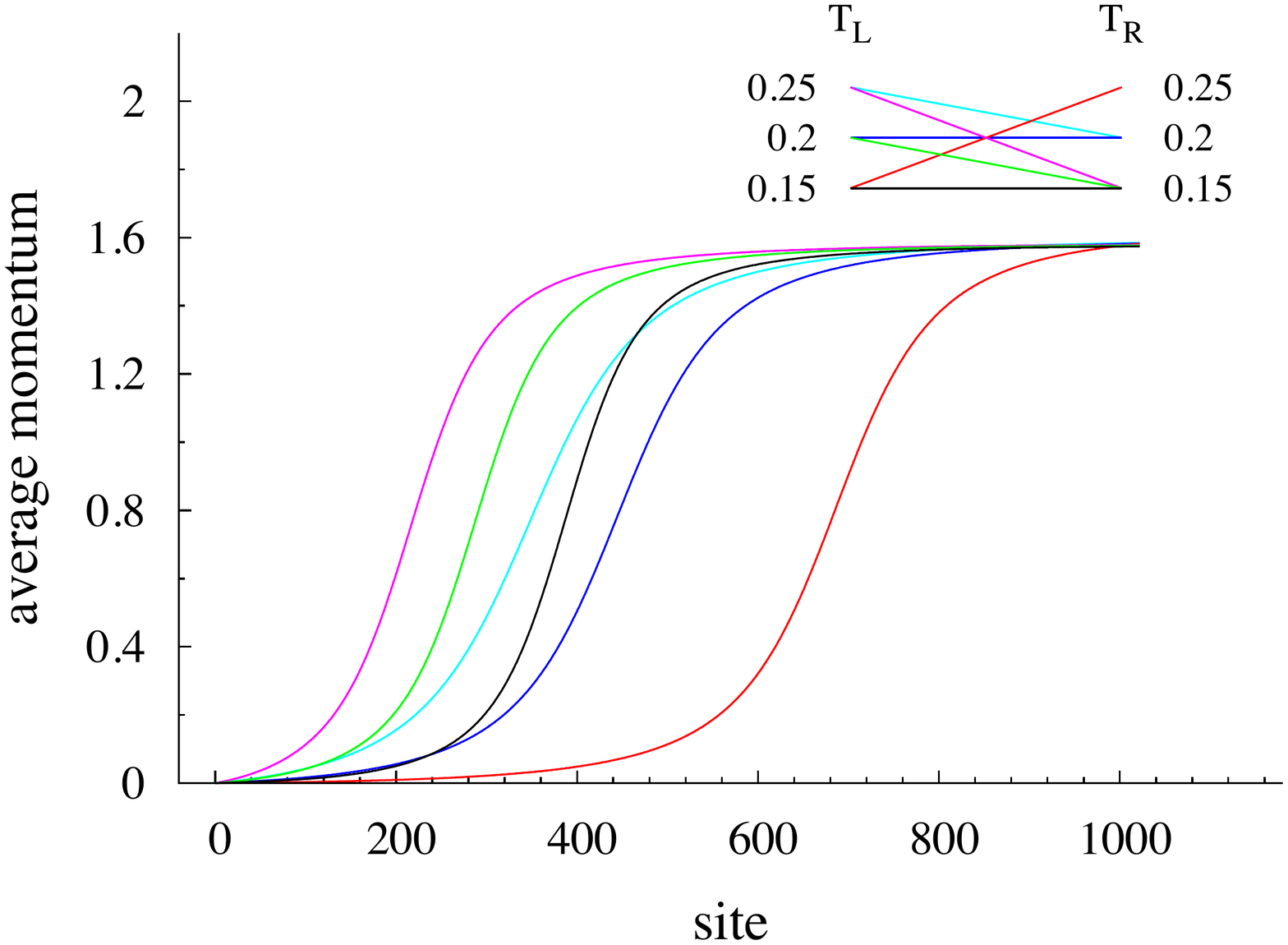}
   \caption{Stationary momentum profiles in rotors dynamics, with $\tau_R = \tau, \tau_L = 0$
       and different values of temperatures for the Langevin heat bath. Reprinted from \cite{iaco}.}
    \label{fig:2}
\end{figure}

  Dynamical simulations of the rotors chain in the stationary state under an exterior torque $\tau$ were
  first perfomed in \cite{iaco}. The resulting stationary profiles of $T_{ss}(\cdot)$ and $p_{ss}(\cdot)$
  are reported in Figures \ref{fig:1} and \ref{fig:2} respectively,
  for different choices of boundary temperatures, and $\tau_L =0$ and $\tau_R = \tau$.
 
  The profiles of temperatures in Figure \ref{fig:1} present a maximum inside the interval 
  with temperatures much higher than at the boundaries. Two flex points are presents
 that get closer as the temperature at the border decreases. 

\begin{figure}[htb]
\centering
  \includegraphics[width=0.60\textwidth]{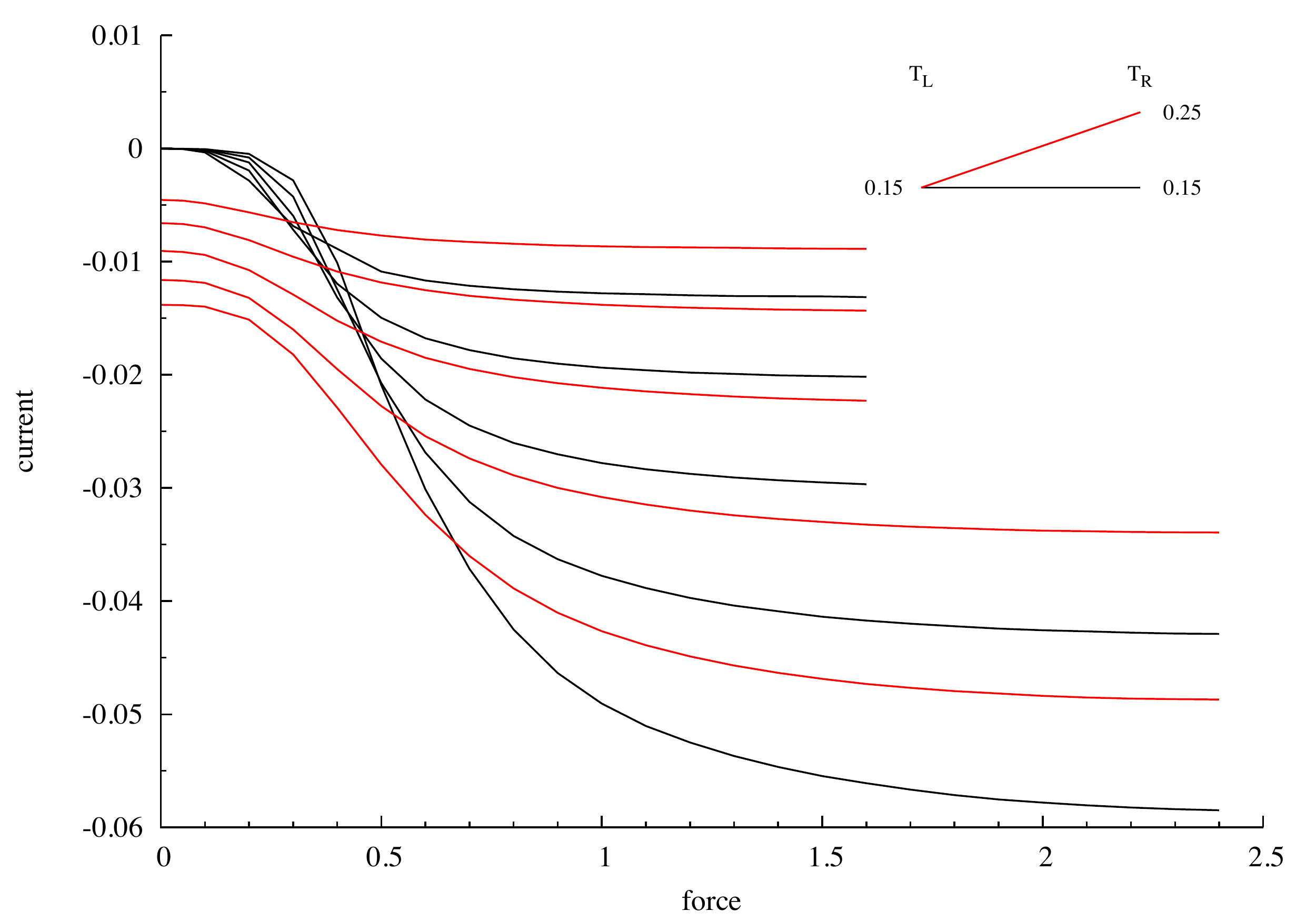}
  \caption{Energy current $J^e$ as function of $\tau_R = \tau$, for $\tau_L = 0$.
    Reprinted from \cite{iaco}.}
    \label{fig:3}
  \end{figure}

  In Figure \ref{fig:2} are the corresponding profiles of $p_{ss}(x)$.
  The maximum of the temperature corresponds to the flex point of $p_{ss}$, in agreement with
  \eqref{eq:pflex}.
  In Figure \ref{fig:3} is the energy current $J^e$ as function of $\tau_R = \tau$ (with $\tau_L = 0$)
  for different sizes of the system $N$. In black is the case of same temperature,
  while in red the curve when $T_R$ is rised.
  Notice in Figure \ref{fig:3} that for $\tau$ large enough the curves
  cross, which implies a negative response to the temperature gradient. This
  is in agreement with the remark made above, as consequence of the decrease of $D^p(T)$ with $T$.

  
  The plots in Figures \ref{fig:1}, \ref{fig:2} and \ref{fig:3}
  are obtained by direct dynamical simulations of the NESS.
  In \cite{iaco2} we attempt to solve numerically equations
  \eqref{eq:3} after having estimated $D^p(T)$ and
  $\kappa(T)$ with dynamical simulations of the system in equilibrium.
  Agreement of the corresponding plots
  will confirm the correctness of the heuristic coming from the linear response theory.

  For a general review about up-hill diffusion, see \cite{krihsna}.
  This phenomenon can also appear in models with phase transitions,
  even if there is only one conserved quantity \cite{cola1}.

  \subsection{Stationary states for harmonic chains with random dynamics}
\label{sec:stat-stat-harm}

\begin{figure}
\begin{center}
\begin{tikzpicture}[scale = 0.85]
\node[circle,fill=black,inner sep=1.2mm] (e) at (0,0) {};
\node[circle,fill=black,inner sep=1.2mm] (f) at (2,0) {};
\node[circle,fill=black,inner sep=1.2mm] (g) at (3.5,0) {};
\node[circle,fill=black,inner sep=1.2mm] (h) at (4.8,0) {};
\node[circle,fill=black,inner sep=1.2mm] (i) at (6.8,0) {};
\node[circle,fill=black,inner sep=1.2mm] (j) at (8.5,0) {};
\node[circle,fill=black,inner sep=1.2mm] (k) at (10,0) {};
\node[circle,fill=black,inner sep=1.2mm] (l) at (11.3,0) {};

\draw[dashed] (2,0) -- (3.5,0);
\draw[dashed] (8.5,0) -- (10,0);
\draw[ultra thick, blue, ->] (11.3,0) -- (12.5,0);

\draw[thick, <->] (3.5,-1.2) -- (4.8,-1.2);

\draw (0, -0.6) node[] {$q_{0}$};
\draw (2, -0.6) node[] {$q_{1}$};
\draw (11.3, -0.6) node[] {$q_{n}$};
\draw (3.5, -0.6) node[] {$q_{x-1}$};
\draw (4.8, -0.6) node[] {$q_{x}$};
\draw (6.8, -0.6) node[] {$q_{x+1}$};
\draw[dashed] (3.5,-1.5) -- (3.5,-1);
\draw (4.1, -1.5) node[] {$r_x$};
\draw[dashed] (4.8,-1.5) -- (4.8,-1);

\draw (-0.6,1.8) node[] {\large\color{red}$T_L$};
\draw (12,1.8) node[] {\large\color{red}$T_R$};
\draw (12.4,-0.4) node[] {\color{blue}$\bar\tau$};

\draw[decoration={aspect=0.3, segment length=3mm, amplitude=3mm,coil},decorate] (0,0) -- (2,0); 
\draw[decoration={aspect=0.3, segment length=1.8mm, amplitude=3mm,coil},decorate] (3.5,0) -- (4.9,0); 
\draw[decoration={aspect=0.3, segment length=3mm, amplitude=3mm,coil},decorate] (4.8,0) -- (6.9,0); 
\draw[decoration={aspect=0.3, segment length=2.5mm, amplitude=3mm,coil},decorate] (6.8,0) -- (8.6,0); 
\draw[decoration={aspect=0.3, segment length=1.8mm, amplitude=3mm,coil},decorate] (10,0) -- (11.4,0); 

\fill [pattern = north east lines, pattern color=red] (-0.3,0.8) rectangle (0.3,2);
\fill [pattern = north east lines, pattern color=red] (11,0.8) rectangle (11.6,2);
\node (c) at (-0.3,1.5) {};
\node (d) at (-0.1,0.1) {};
\node (a) at (11.6,1.5) {};
\node (b) at (11.4,0.1) {};

\draw (c) edge[dashed, ultra thick, red, ->, >=latex, bend right=60] (d);
\draw (a) edge[dashed, ultra thick, red, ->, >=latex, bend left=60] (b);

\end{tikzpicture}
\end{center}
\label{fig:harm}
\caption{Chain of harmonic oscillators with random flip pf velocities sign,
  heat bath and tension applied on teh right hand side. Reprinted from \cite{kos2}. }
\end{figure}
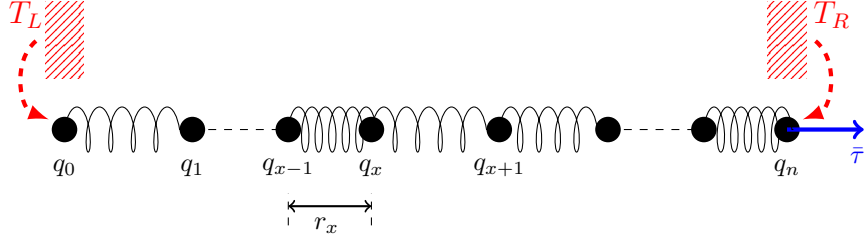

  Mathematical rigorous results can be obtained for the NESS of the harmonic oscillators dynamics
  with random exchange of velocities between nearest neighbor particles,
  such that kinetic energy is conserved. The conserved quantities are the volume 
  and the total energy, and the non-stationary evolution is governed
  in the diffusive scaling by \eqref{eq:20}.
  In \cite{kos2} we study this dynamics when the system has $N$ harmonic oscillators connecting $N+1$ particles,
  with Langevin heat baths
  attached at the first left particle and the last right particle, and a constant force $\tau$
  is attached to the last right particle (see Figure \ref{fig:harm}).
  The system is driven out of equilibrium by the presence of the external force $\tau$,
  and the stochastic part of the dynamics has only equilibrium states with $0$ average velocities. 
  Thanks to the stochastic dynamics in the bulk, the NESS exists.
  For $\tau = 0$ this NESS was studied in \cite{bo1}.
  
  In \cite{kos2} we prove the hydrodynamic limit in the NESS, and the stationary profiles
  of volume stretch $r_{ss}(x)$ and temperature $T_{ss}(x)$ satisfy the equations
  \begin{equation}
    \label{eq:26}
    \partial_x^2 r_{ss}(x) = 0, \qquad \partial_x^2 T_{ss}(x)= -2 \left(\partial_x r_{ss}(x)\right)^2,
    \qquad r_{ss}(0) = 0, \ T_{ss}(0) = T_L, \quad r_{ss}(1) = \tau, \ T_{ss}(1) = T_R.
  \end{equation}
  These can be explicitly solved obtaining $r_{ss}(x) = \tau x$ and
\begin{equation*}
   T_{ss}(x) = \tau^2 x (1-x) + (T_R-T_L)x + T_L,  \qquad x\in [0,1],
 \end{equation*}
 that shows again an heating phenomena at the center of the system.
 The stationary energy current can also be calculated and gives
 \begin{equation}
   \label{eq:27}
   J^e = -\frac 1{4\gamma} \left(T_R - T_L\right) -\frac{\tau^2}{2\gamma}.
 \end{equation}
 So if uphill diffusion is possible, no negative response to temperature gradient
 can happen in this system,
 as these were due to the non-linearities in the temperature of the diffusivities in the rotor chain.
 Also there are noflex points in these profiles of temperature, as also these were due
 to the temperature dependence of the thermal conductivity.
  
\section{The two steps approach:
  weak coupling limits, kinetic limits and hydrodynamic limits}
\label{sec:two-step-approach}

As we mentioned in Section \ref{sec:diff-macr-equat}, one of the major difficulties in order to obtain the diffusive equations
in the hydrodynamic limit is that it involves a simultaneous scaling of space and time.
One way to simplify the problem is to introduce a small parameter in the dynamics that makes the interaction
weak (\emph{weak coupling}) or the collisions rarefied (\emph{kinetic limit}), in order to break the procedure in two steps:
\begin{itemize}
\item a weak coupling or kinetic limit
  where it is obtained an autonomous \emph{mesoscopic} stochastic dynamics,
  \item a subsequent hydrodynamic limit for these stochastic dynamics.
\end{itemize}

\subsection{Weak coupling limit }
\label{sec:weak-coupling-limit}

In the case of dynamics that conserve macroscopically only energy,
some progress have been obtained in the past years in the weak coupling limit,
even though the full program is not yet complete.

Consider the generic Hamiltonian dynamics
\begin{equation}
  \label{eq:2-gen}
    \dot q_i(t) = p_i(t) , \qquad \dot p_i (t) = \delta V'(q_{i+1}(t) - q_i(t)) - \delta V'(q_{i}(t) - q_{i-1}(t)) + U'(q_i(t)),
  \end{equation}
  where $\delta>0$ is a small parameter,  $q_i\in M$ where $M$ is some d-dimensional manifold,
  and $U$ is a potential
  defined on $M$. With $q_{i+1}- q_i$ is intended some distance defined on $M$.
  When $\delta = 0$, there will be no
  interaction between the particles and no exchange of energy. When $\delta >0$ there will be an
  instantaneous energy current between the particles given by
  \begin{equation}
    \label{eq:28}
   \delta j^e_{i,i+1}(t) = -\delta p_i(t)\cdot V'(q_{i+1}(t)- q_i(t)).
  \end{equation}
  Notice that in any equilibrium the average of $j^e$ is null.
  In order to see a energy diffusion in the limit as $\delta \to 0$,
  we have to look at a time scale $\delta^{-2} t$. In fact
  \begin{equation}
    \label{eq:29}
   e_i (\delta^{-2}t) = \delta \int_0^{\delta^{-2}t}  ( j^e_{i-1,i}(s) - j^e_{i,i+1}(s))\; ds
 \end{equation}
 Let us assume that a central limit theorem is valid for the uncoupled dynamics ($\delta = 0$),
 and \emph{somehow} stable for small $\delta>0$.
 The first step consists in proving that $e_i (\delta^{-2}t)$ converges in law
 to an autonomous stochastic dynamics of energies:
 \begin{equation}
   \label{eq:30}
   e_i (\delta^{-2}t) \ \mathop{\longrightarrow}_{\delta\to 0}\ \mathcal E_i(t),
 \end{equation}
 where the $\mathcal E_i(t)$ satisfy the stochastic differential equations
 \begin{equation}
   \label{eq:31}
   \begin{split}
     d \mathcal E_i(t) &= dJ_{i-1, i}(t) - dJ_{i, i+1}(t), \\
     dJ_{i, i+1}(t) &= \alpha(\mathcal E_i(t), \mathcal E_{i+1}(t)) \; dt +
     \sqrt 2 \sigma^2(\mathcal E_i(t), \mathcal E_{i+1}(t)) \; dB_i(t),
   \end{split}
 \end{equation}
 where $B_i(t)$ are independent standard Wiener processes, and $\sigma^2$ are the variances
 of the energy currents $j^e_{i,i+1}(t)$ in the CLT of the uncoupled dynamics:
 \begin{equation}
   \label{eq:32}
   \sigma^2(\mathcal E_1, \mathcal E_{2}) \ =
   \ \int_0^\infty \left<  j^e_{1,2}(t)  j^e_{1,2}(0)\right>_{\mathcal E_1, \mathcal E_{2}} \ dt.
 \end{equation}
 Here $ \left<  \cdot \right>_{\mathcal E_1, \mathcal E_{2}}$ denotes the expectation with respect to the uncoupled dynamics
 of two particles in the microcanonical equilibrium at fixed energies $\mathcal E_1$ and $\mathcal E_{2}$.
 The functions $\alpha(\mathcal E_1, \mathcal E_{2})$ are antisymmetric and are defined by
 \begin{equation}
   \label{eq:33}
   \alpha(\mathcal E_1, \mathcal E_{2}) =
   \left( \partial_{\mathcal E_1} - \partial_{\mathcal E_2}\right) \sigma^2(\mathcal E_1, \mathcal E_{2})
   + \sigma^2(\mathcal E_1, \mathcal E_{2}) \left( \frac{Z'(\mathcal E_1)}{Z(\mathcal E_1)}
     - \frac{Z'(\mathcal E_2)}{Z(\mathcal E_2)}\right),
 \end{equation}
 where $Z(\mathcal E)$ is the volume on the microcanonical manifold of energy $\mathcal E$ of
 the single uncoupled particle. The equations \eqref{eq:32} define a stochastic dynamics reversible with respect to
 the stationary measures:
 \begin{equation}
   \label{eq:36}
   d\tilde \nu_\beta = \prod_i \frac{Z(\mathcal E_i) e^{-\beta \mathcal E_i}}{\tilde Z_\beta} \; d\mathcal E_i 
 \end{equation}

 A proof of this first step, i.e. the limit \eqref{eq:30}, would require that the uncoupled dynamics
 is \emph{chaotic} enough such that a
 CLT theorem is valid and is \emph{stable} for small perturbations.

 In \cite{DL} this is proven for
 particles moving (deterministically) as geodesic flow in a manifold $M$ with strictly negative curvature,
 with dynamical system techniques.
 For anharmonic oscillators with a stochastic noise acting on the velocities and conserving energy,
 the first step has been proven in \cite{LO}, using hypocoercive estimates. One particular case of \cite{LO} is the
 harmonic case ($V$ and $U$ quadratic), where
 $\sigma^2(\mathcal E_1, \mathcal E_{2}) = \gamma^{-2} \mathcal E_1 \mathcal E_{2}$, and
 $\alpha(\mathcal E_1, \mathcal E_{2}) = \gamma^{-2} (\mathcal E_1 - \mathcal E_{2})$,
 where $\gamma$ is the intensity of the noise in the dynamics.

 The second step consists in obtaining the hydrodynamic limit for the energy evolution
 of the stochastic dynamics given by \eqref{eq:31}, i.e. that we have the convergence of the empirical distributions
 \begin{equation}
   \label{eq:34}
   \epsilon \sum_i \delta_{\epsilon i}(dx) \mathcal E_i(\epsilon^{-2} t) \ \mathop{\longrightarrow}_{\epsilon\to 0}
   \ \mathcal E(t,x) dx,
 \end{equation}
where
\begin{equation}
  \label{eq:35}
  \partial_t \mathcal E = \partial_x \left(\tilde D(\mathcal E) \partial_x \mathcal E\right),
\end{equation}
with the energy diffusivity of the stochastic dynamics given by $\tilde D = \tilde C_v^{-1} \tilde \kappa$,
where $\tilde C_v$ is the variance of the energies $\mathcal E_i$ under  $d\tilde \nu_\beta$
and the conductivity $\tilde \kappa$ is given by the corresponding Green-Kubo formula
\begin{equation}
  \label{eq:37}
  \tilde \kappa (\beta^{-1})  = \beta^2 \sum_i \int_0^\infty \left<  \alpha(\mathcal E_i(t), \mathcal E_{i+1}(t)) ,
    \alpha(\mathcal E_0(0), \mathcal E_{1}(0)) \right>_\beta \; dt.
\end{equation}
The relation between  $\tilde \kappa (T)$ and the thermal conductivity of the original dynamics with $\delta >0$,
i.e. $\kappa_\delta(T)$ defined as in \eqref{eq:14}, is studied in \cite{bhllo}, and it turns out that
\begin{equation}
  \label{eq:38}
  \kappa_\delta(T) = \tilde \kappa (T) \delta + o(\delta).
\end{equation}

The reversible stochastic dynamics \eqref{eq:31} is a version of the conservative Ginzburg-Landau dynamics
considered by Varadhan in \cite{Var}. With respect to \cite{Var}, this dynamics is restricted on $\mathbb R_+^{\mathbb Z}$,
and we call the \emph{energy Ginzburg-Landau dynamics}. In order to apply the method of \cite{Var}
we need a lower bound 
on the spectral gap for the generator on the corresponding finite dimensional dynamics, i.e. it should be bounded below
by $CN^{-2}$, where $N$ is the dimension of the system, for some constant $C$
than may depend on the energy but not on the dimension.
When this spectral gap bound can be proven, the second step can be perfomed \cite{LOS}. This is the case
for the \eqref{eq:31} arising from the weak coupling limit in the anharmonic chain with noise (cf. \cite{LO})
with some conditions on the pinning potential $U$. In fact if $\text{Hess}\; U(0)> 0$, it follows that
$\sigma^2 (\mathcal E_1, \mathcal E_{2}) = \mathcal E_1 \mathcal E_{2} G(\mathcal E_1, \mathcal E_{2})$,
with $G\ge c>0$.
This implies that the generator of \eqref{eq:31} has a spectral gap bound (this can be proven following the
argument in \cite{os}).
For the energy Ginzburg Landau dynamics emerging from the deterministic dynamics of the geodesic flows
\cite{DL}, we have that
\begin{equation}\label{eq:dlm}
\sigma^2 (\mathcal E_1, \mathcal E_{2}) \ \sim\ 
\frac{\mathcal E_1 \mathcal E_{2}}{ \mathcal E_1^{3/2}+ \mathcal E_{2}^{3/2}}.
\end{equation}
Unfortunately under a behaviour like \eqref{eq:dlm} it is not clear that a spectral gap bound
will hold, it certainly depends on the energy. At
 this time the step 2 remains an open problem when starting from a purely Hamiltonian deterministic
dynamics.

\subsection{Kinetic limit}
\label{sec:kinetic-limit}

A different two step approach consists in studying models where energy is exchanged between particles through
\emph{collisions} that are rarefied because of constraints in the geometry of the system. Time is scaled
in such way that on a unit time there is, in average, a finite number of collisions per unit time. In this sense the first step
is similar to the Boltzmann-Grad limit.

A typical model considered in this approach (cf. \cite{blps92} \cite{gl08})
is given by a chain of Sinai's billiards, where each particle is
confined in one billiard, but it can collide with a particle in a neighbour billiard through a small window.
The size $\delta$ of this window is the small parameter of the limit. The energy is the only conserved quantity, and
it is exchanged through these collision. Because of the chaoticity of the billiard, after time rescaling as $\delta\to 0$,
the difference in the collision times, conditioned on the energy of each billiard, became independent. Consequently
it is expected that the energies per particle $e_i(\delta^{-1} t)$ converge to a Markov jump process
$\mathcal E_i(t)$ whose generator is given by
\begin{equation}
  \label{eq:39}
  \begin{split}
    L F(\{\mathcal E_j\}) &=  \sum_i  \int_0^1 d\alpha \Lambda(\mathcal E_i, \mathcal E_{i+1}, \alpha)
    \left[ F(T_{i,i+1,\alpha} \{\mathcal E_j\}) - F( \{\mathcal E_j\})\right]\\
  (T_{i,i+1,\alpha} \{\mathcal E_j\})_k &= \alpha ( \mathcal E_i + \mathcal E_{i+1}) \delta_{(k=i)}
  + (1-\alpha)  ( \mathcal E_i + \mathcal E_{i+1}) \delta_{(k=i+1)} +  \mathcal E_k \delta_{(k\neq i,i+1)}.
\end{split}
\end{equation}
This means that at random times, exponentially distributed with intensity
$ \Lambda(\mathcal E_i, \mathcal E_{i+1}, \alpha)$,
the total energy of the site $i$ and $i+1$ is redistributed in the two sites with proportion $\alpha$ and $1-\alpha$.
The precise form of the rate function $\Lambda$ can be found in \cite{gl08} and \cite{makikosg},
but what matters is that $\Lambda(\mathcal E_i, \mathcal E_{i+1}, \alpha) \sim ( \mathcal E_i + \mathcal E_{i+1})^{1/2}$.

Unlike the weak coupling limit, no rigorous results about this first step exist at the moment starting from the deterministic dynamics. Some partial attemps and a detailed discussion of the problem can be found in
\cite{hung-17}.

The second step will be the hydrodynamic limit on the stochastic dynamics generated by \eqref{eq:39}
in order to obtain the diffusive equation for the energy.
This is a \emph{non-gradient} dynamics where in principle could be applied Varadhan's approach \cite{Var}.
With respect to the stochastic dynamics emerging in the weak coupling case from a purely mechanical model, there
exists a proof of the spectral gap bound (cf. \cite{makikosg}), necessary in order to apply Varadhan's method.
Still there are other intrinsic difficulties that did not allow yet to prove the hydrodynamic limit.



\section*{Acknowledgements}
This article contains ideas and results developped in collaboration with Cedric Bernardin, Fran\c cois Huveneers,
Alessandra Iacobucci, Tomasz Komorowski, Joel Lebowitz, Carlangelo Liverani, Makiko Sasada, 
Marielle Simon, Gabriel Stoltz.

This work is supported by the grant ANR-15-CE40-0020-01 LSD 
of the French National Research Agency.


\begin{thebibliography}{00}

  
\bibitem{hung-17} Bálint, Péter; Gilbert, Thomas; Nándori, Péter; Szász, Domokos; Tóth, Imre Péter On the limiting Markov process of energy exchanges in a rarely interacting ball-piston gas. J. Stat. Phys. 166 (2017), no. 3-4, 903–925.

\bibitem{ber05} C. Bernardin,
  Hydrodynamics for a system of harmonic oscillators perturbed by a conservative noise.
\emph{Stoc. Proc. and Appl.}, 117 (2007) 487–513.
  
\bibitem{bhllo} C. Bernardin, F. Huveneers, J. L. Lebowitz, C. Liverani, S. Olla,
  Green-Kubo Formula for Weakly Coupled Systems with Noise,
  Commun. Math. Phys., 334, n.3, 1377–1412, March 2015. DOI: 10.1007/s00220-014-2206-7
  
\bibitem{bo1} C. Bernardin, S. Olla, Transport
Fourier’s Law for a Microscopic Model of Heat Conduction, 
{\it{J. Stat. Phys.}}, 121: 271--289, 2005.   


\bibitem{bo2} C. Bernardin, S. Olla, Transport
Properties of a Chain of Anharmonic Oscillators with Random Flip of
Velocities, {\it{J. Stat. Phys.}}, 145: 1224--1255, 2011.   

\bibitem{blps92} Bunimovich, L., Liverani, C., Pellegrinotti, A., Suhov, Y. (1992).
  Ergodic systems of n balls in a billiard table. \emph{Comm.  Math. Phys.}, 146(2), 357-396.
URL http://dx.doi.org/10.1007/BF02102633

\bibitem{cola1} Colangeli, M. , De Masi, A. , Presutti, E.:
  Microscopic models for uphill diffusion. \emph{J.Phys.A}, 50, 435002, (2017).

\bibitem{cuneo2} N. Cuneo, J.-P. Eckmann, and C. Poquet.
  Non-equilibrium steady state and subgeometric ergodicity
  for a chain of three coupled rotors. \textit{Nonlinearity}, 28(7):2397–2421, 2015.

  
\bibitem{cuneo1} N. Cuneo and C. Poquet.
  On the relaxation rate of short chains of rotors interacting with Langevin thermostats.
  \emph{Electronic Communications in Probability}, 22:8 pp., 2017.

\bibitem{ddd} A. Das, K. Damle, A. Dhar, D. A. Huse, M. Kulkarni, C. B. Mendl and H. Spohn,
  Nonlinear Fluctuating Hydrodynamics for the Classical XXZ Spin Chain, arXiv:1901.00024v1 (2019).


\bibitem{das-dhar}  Suman G. Das and A. Dhar. Role of conserved quantities in normal heat transport in one dimenison. arXiv preprint arXiv:1411.5247, 2014.

\bibitem{DL} D. Dolgopyat and C. Liverani, 
Energy transfer in a fast-slow Hamiltonian system,
\textit{Commun. Math. Phys.} {\bf 308}, 201--225 (2011).

\bibitem{ffl} Fritz, J.,  Funaki, T., Lebowitz, J. L., Stationary
  states of random Hamiltonian systems.  {\it{Probab. Theory Related
      Fields}}, 99(2):211--236, 1994.


 \bibitem{gl08} P. Gaspard, T. Gilbert,
   Physical Review Letters 101(2), 20601 (2008). http://dx. doi.org/10.1103/PhysRevLett.101.020601

\bibitem{iaco} A. Iacobucci, F. Legoll, S. Olla, G. Stoltz, Negative thermal conductivity of chains of rotors with
  mechanical forcing, \textit{Phys. Rev. E}, 84, 061108, 2011.

  
\bibitem{iaco2} A. Iacobucci, S. Olla, G. Stoltz, Stationary non-equilibrium states in rotors models, in preparation.

\bibitem{illp14} S. Iubini,  S. Lepri, R. Livi, A. Politi, {Boundary induced instabilities in coupled
    oscillators}, \textit{Phys. Rev. Lett.} 112, 134101, 2014.

\bibitem{illp16} S. Iubini S. Lepri, R. Livi, A. Politi, Coupled transport in rotor models, \emph{New Journal of Physics},
  2016.  (https://arxiv.org/pdf/1603.06847.pdf)
  
 \bibitem{jko} M. Jara, T. Komorowski, S. Olla, {Superdiffusion
    of Energy in a system of harmonic oscillators with noise}, 
\textit{Commun. Math. Phys.}  339: 407, 2015.


\bibitem{kipnis} Kipnis, C. and Landim, C.,
{\it Scaling Limits of Interacting Particle Systems},
Springer-Verlag: Berlin, 1999.

\bibitem{kona}  T. Komorowski, S. Olla, Diffusive propagation of energy in a non-acoustic chain,
  \textit{Arch. Rat. Mech. Appl.} 223, N.1, 95--139, 2017.
 
\bibitem{kos1}  T. Komorowski, S. Olla, M. Simon,
  Macroscopic evolution of mechanical and thermal energy in a harmonic chain
  with random flip of velocities, \textit{Kinetic and Related Models}, AIMS, 11 (3):  615--645, 2018.

\bibitem{kos2}  T. Komorowski, S. Olla, M. Simon,
  An open microscopic model of heat conduction: evolution and non-equilibrium stationary states, 
  http://arxiv.org/abs/1903.11374, 2019.
  
\bibitem{krihsna} R. Krishna, Uphill diffusion in multicomponent mixtures, \textit{Chem. Soc. Rev.}, 44, 2812--2836, 2015.

  
  
  
  
\bibitem{LO} C. Liverani, S. Olla,  Toward the Fourier law for a weakly interacting anharmonic crystal,
  \textit{JAMS} {\bf 25}, N. 2, 555--583, (2012).

\bibitem{LOS} C. Liverani, S. Olla,  M. Sasada, Diffusive scaling in energy Ginzburg-Landau dynamics, 2015,
  http://arxiv.org/abs/1509.06116.

  \bibitem{os} S. Olla, M. Sasada, Macroscopic energy diffusion for a
  chain of anharmonic oscillators, \textit{Probab. Theory Relat. Fields}, 
 \textbf{157},
  721--775 (2013), DOI 10.1007/s00440-012-0469-5.


\bibitem{so1} S. Olla, M. Simon, Microscopic derivation of an adiabatic thermodynamic transformation,
  \emph{Brazilian Journal of Probability and Statistics}, 2015, Vol. 29, No. 2, 540–564, 
  DOI: 10.1214/14-BJPS275

\bibitem{so2} S. Olla, M. Simon, in preparation.
  
  

  
\bibitem{quastel} J. Quastel, Diffusion of colors in the simple exclusion process,
  \textit{Comm. Pure Appl. Math.} 45, N. 6, (1992).

\bibitem{makikosg} M. Sasada, Spectral gap for stochastic energy exchange model  with non-uniformly positive rate
  function, \emph{Ann.Prob.}, 2015, 43, n.4, 1663–1711, DOI: 10.1214/14-AOP916.  
  

  
\bibitem{spohn-rot}  Herbert Spohn. Fluctuating hydrodynamics for a chain of nonlinearly coupled rotators.
  arXiv:1411.3907, 2014.
  
\bibitem{Var} S.R.S. Varadhan, 
Nonlinear diffusion limit for a system with nearest neighbor
interactions II, \textit{Asymptotic problems in probability theory: stochastic
models and diffusions on fractals} (Sanda/Kyoto, 1990), 75-128, \textit{ Pitman
Res. Notes Math. Ser.}, {\bf 283}, Longman Sci. Tech., Harlow, (1993).   
\end{thebibliography}
\end{document}